\documentclass[preprint,12pt]{elsarticle}

\usepackage{amssymb}
\usepackage{graphicx}
\usepackage{verbatim}  
\usepackage{marvosym}  
\usepackage{amsmath}
\usepackage{amssymb}
\usepackage{euscript}
\usepackage{wasysym}
\usepackage{mathrsfs}
\usepackage{bm}
\usepackage{setspace} 
\usepackage{color}
\usepackage{amsfonts}

\newcommand{\addfigure}[5][h]
{
	\begin{figure}[#1]
		\begin{center}
			\includegraphics[#5]{#2}
			\caption{#3.\label{#4}}
		\end{center}
	\end{figure}
}

\begin{document}

\begin{frontmatter}


\cortext[cor1]{a.iliyasu@psau.edu.sa}

\title{Circuit-based modular implementation of quantum ghost imaging}

 \author[label1]{Fei Yan}
 \author[label1]{Kehan Chen}
 \author[label1,label2]{Abdullah M. Iliyasu\corref{cor1}}
 \author[label1]{Jianping Zhao}
 \address[label1]{School of Computer Science and Technology, Changchun University of Science and Technology, Changchun 130022, China}
 \address[label2]{Electrical Engineering Department, College of Engineering, Prince Sattam Bin Abdulaziz University, Al-Kharj 11942, KSA}

\begin{abstract}
Efforts on enhancing the ghost imaging speed and quality are intensified when the debate around the nature of ghost imaging (quantum vs. classical) is suspended for a while. Accordingly, most of the studies these years in the field fall into the improvement regarding these two targets by utilizing the different imaging mediums. Nevertheless, back to the raging debate occurred but with different focus, to overcome the inherent difficulties in the classical imaging domain, if we are able to utilize the superiority that quantum information science offers us, the ghost imaging experiment may be implemented more practically. In this study, a quantum circuit implementation of ghost imaging experiment is proposed, where the speckle patterns and phase mask are encoded by utilizing the quantum representation of images. To do this, we formulated several quantum models, i.e. quantum accumulator, quantum multiplier, and quantum divider. We believe this study will provide a new impetus to explore the implementation of ghost imaging using quantum computing resources.

\end{abstract}


\end{frontmatter}



\section{Introduction}
\label{sec1}
Leveraging on its immense potentials for applications that require minimal computing resources, speed, security, etc., quantum information science has exploded beyond its utility in optics (including laser technology \cite{Kazakov2017Self} and remote sensing technology \cite{Takenaka2017Satellite}) to exciting applications in computer science and engineering (e.g., machine learning \cite{Hush2017Machine} and artificial intelligence \cite{Trabesinger2017Quantum}). Naturally, this has also led to inter-disciplinary explorations, such as quantum ghost imaging \cite{Boyd2012Introduction} and quantum image processing \cite{Venegas2015Introductory}, etc.

Ghost imaging is a technique employed to retrieve an object from the cross-correlation function of two separate beams and neither of which obtains the information from the object \cite{Basano2007A}. One beam interrogates a target and then illuminates a single-pixel detector that provides no spatial resolution, while the other beam does not interact with the target, but it impinges on a high-resolution camera, hence affording a multiple-pixel output \cite{erkmen2010ghost}. The timeline for this sub-discipline's
development shows its modest beginning started in 1995, where the two beams of ghost imaging were formed from a stream of entangled photons \cite{Pittman1995Optical}. The reconstruction of the image was attributed to the non-local quantum correlations between the photon pairs. For several years, the ghost imaging was considered as an effect of quantum non-locality due to the earlier experiments. Challenging this interpretation, Bennink et al. demonstrated ghost imaging using two classically correlated beams \cite{Bennink2002Two}, following which, it was found that many of the features obtained with entangled photons could be reproduced with a classical pseudothermal light source. However, the nature of the spatial correlations exhibited with a pseudothermal source, and whether they can be interpreted as classical intensity correlations or are fundamentally non-local quantum correlations, is still under debate \cite{Gatti2007Comment,Scarcelli2007Scarcelli,Shih2012the,Shapiro2012the}. Although, focusing on the problems and improvement of the spatial resolution, field of view, and signal-to-noise ratio of the ghost imaging result, researchers have gained a lot of progresses by using different types of light sources \cite{Schneider2018quantum} and the implementations on different materials \cite{Tetienne2017quantum}.

Even though the ghost imaging technique has shown potential for applications demanding high detection sensitivity as well as high resolution, which are useful in the civil and military domains, the following reasons impede its development to a large extent:
\begin{itemize}
  \item To assure the quality of the ghost imaging result, a large amount of samplings for speckle patterns are required. Establishment (i.e. representation) and storage of these patterns will be a heavy job for classical computers.
  \item In order to generate the interested image, the interactions between the speckle patterns and the phase mask require a huge number of judgement (i.e. judging whether the pixel in the speckle patterns is in the subregion of the phase mask) which will cost a massive computing resources.
  \item Cross correlation of the signals in the signal and idler fields will also undertake a great deal of computations which is also a great trials to the classical computing devices.
\end{itemize}

Fortunately, these problems occurred on classical computing domain may be solved by utilizing the quantum computing framework \cite{Feynman1982Simulating,Shor1994Algorithms,Deutsch1985Quantum}. Among many other areas, these tools are used in the emerging sub-discipline of quantum image processing. Technically, quantum image processing is focused on extending conventional image processing tasks and operations to the quantum computing framework \cite{iliyasu2013towards,Aburaed2017Advances,yan2017qip}. It is primarily devoted to utilizing quantum computing technologies to capture, manipulate, and recover quantum images in different format and for different purposes \cite{yan2016asurvey}. Due to some astounding properties of quantum computation (i.e. entanglement and parallelism), it is anticipated that quantum image processing will offer incredible capabilities and performances in terms of computing speed, guaranteed security, and minimal storage requirements, etc. \cite{yan2016asurvey}. The pioneering work of quantum image processing should be attributed to Venegas-Andraca and Bose's Qubit Lattice \cite{venegas2003storing} description for quantum images in 2003, while the proposal of flexible representation for quantum images \cite{le2011a} by Le et al. in 2010 receives more attention since it supports the integration of the quantum image into a normalized state and facilitates auxiliary transformations on the content of the image. Following these, many other quantum image representations have been proposed \cite{jiang2015quantum,li2014multi} as well as an array of algorithmic frameworks that target the spatial or chromatic content of the image \cite{Yao2017quantum,caraiman2013histogram,iliyasu2012watermarking,zhou2017quantum}.

In this study, we attempt to utilize the proven potency of quantum information processing, i.e. quantum image processing, in a new paradigm for ghost imaging. The advantages are provided as follows:
\begin{itemize}
  \item Quantum register which includes \emph{n} quits is able to store $2^n$ binary numbers. Such exponential storage ability (in comparison with the classical register/storage) on quantum computers could solve the problem that large amounts of speckle patterns occupy too much storage space commendably.
  \item The judgements required in the interaction between the speckle patterns and the phase mask could be simply done by several quantum CNOT gates. Utilizing the parallelism of quantum computing, the complexity of the whole simulating calculation would be greatly reduced.
  \item Cross correlation at the final step would also benefit from the parallelism of quantum computing. By designing the quantum arithmetic operations, i.e. quantum accumulator, quantum multiplier, and quantum divider, the efficient operations (and circuit implementation) of the cross correlations could be assured.
\end{itemize}

The rest of the paper is organized as follows: in Section \ref{sec2}, quantum image representation, quantum adder, and quantum comparator are introduced, following these, the quantum accumulator, quantum multiplier, and quantum divider are designed and proposed. In Section \ref{sec3}, a complete quantum ghost imaging circuit is established including the creation of quantum speckle patterns, the interaction between the patterns and the quantum phase mask, and quantum computation of the cross-correlation. The conclusions of the study and its applications on the quantum image encryption are discussed in Section \ref{sec5}.

\section{Modular approach to basic quantum arithmetic operations}
\label{sec2}
Traditionally, the arithmetic operation of addition, subtraction, multiplication, and division are employed to operation on two or more numbers. As quantum states, it is infeasible to extend classical execution of these operations as natural numbers to our quantum ghost imaging (QGI) framework. In addition to the four traditional operations, we utilize the accumulator (ACC) and comparator (COM) operations as the quantum arithmetic operations to support the execution of our proposed QGI protocol. First, we formalize, from established literature, the notion of an image as used on the quantum computing paradigm.

\subsection{Quantum image representation}
\label{subsec2-1}
As defined in \cite{iliyasu2013towards}, quantum image processing is devoted to ``utilizing the quantum computing technologies to capture, manipulate, and recover quantum images in different formats and for different purposes.'' The first step accomplishing this requires representation to encode images based on the quantum mechanical composition of any potential quantum computing hardware be conjured \cite{yan2016a}. Among all of the available quantum image representations, in this study, the novel enhanced quantum representation of digital images \cite{Zhang2013NEQR} is utilized to represent the speckle pattern and the target image, which supports the use of the basis states of two qubit sequence to store the chromatic and spatial content each pixel in the image and mathematically defined as follows:

\begin{equation}\label{eq1}
\begin{aligned}
|I(m,n,p)\rangle &= {1\over 2^{(m+n)/2}}\sum_{y=0}^{2^m-1}\sum_{x=0}^{2^n-1}\bigotimes_{h=0}^{l-1}|c_{yx}^h\rangle|yx\rangle,
\end{aligned}
\end{equation}
where $|c_{yx}^h\rangle$ ($c_{yx}^h\in\{0,1\}$) encodes the chromatic information of the pixel at position $|yx\rangle$, where $|yx\rangle=|y\rangle|x\rangle=|y_{m-1}y_{m-2}\ldots y_0\rangle|x_{n-1}x_{n-2}\ldots x_0\rangle$. An example of a $2\times 2$ quantum image and its quantum state is presented in \cite{yan2016asurvey}, wherein, its preparation and retrieval procedures have been thoroughly discussed.

\subsection{Quantum adder}
\label{subsec2-2}
\begin{figure}[!htb]
  \centering
  \includegraphics[width=\hsize]{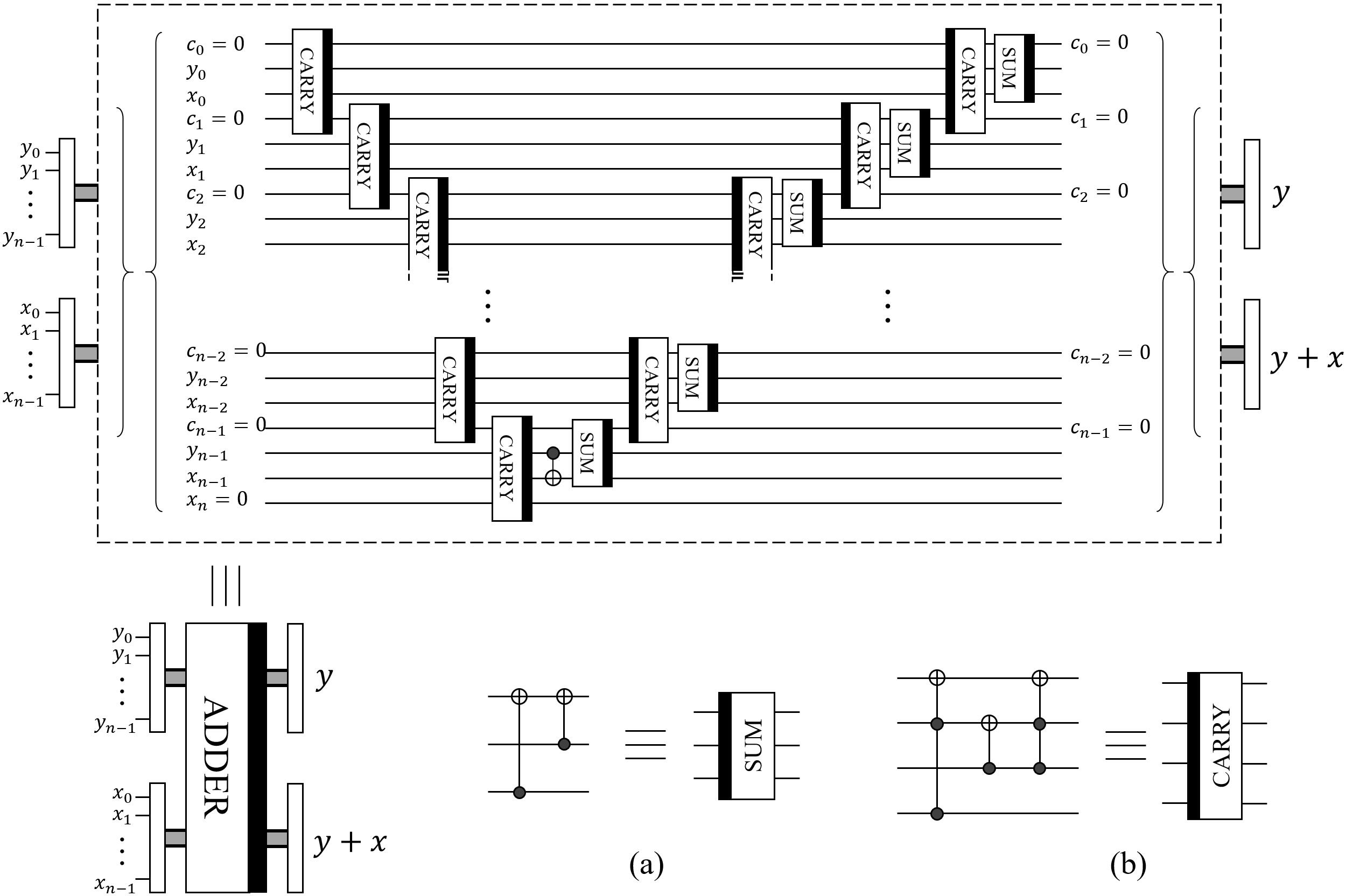}
  \caption{Circuit implementation of quantum ADD module (figure and descriptions adapted from \cite{yan2017quantum})}
  \label{fig1}
\end{figure}

Quantum adder (hereinafter called ADD module) is considered a basic quantum arithmetic operation in the quantum computing field \cite{Vedral1996Quantum}. The aim of ADD module is to perform the following equation:
\begin{equation}\label{eq2}
\text{ADD}\vert y,x\rangle=\vert y, y+x\rangle,
\end{equation}
where $|y\rangle $ and $|x\rangle$ are two input quantum kets and the two output kets are $|y\rangle $ and $|y+x\rangle$. As presented in Figure \ref{fig1}, a quantum adder consists of $2n-1$ carry modules and $2n$ sum modules. In addition, the carry module could be decomposed to 2 Toffoli gates and 1 CNOT gate, while the sum module could be executed by 2 CNOT gates as presented in Figure \ref{fig1}(a) and (b). Moreover, as discussed in \cite{Vedral1996Quantum} and \cite{Draper2000Addition}, quantum subtraction could be implemented using a network of quantum adder(s) due to the fact that quantum gates are reversible. To illustrate the subtracter (i.e. SUB), a black bar is inserted on the left side of the module block.

\subsection{Quantum comparator}
\label{subsec2-c}
\begin{figure}[!htb]
  \centering
  \includegraphics[width=10 cm]{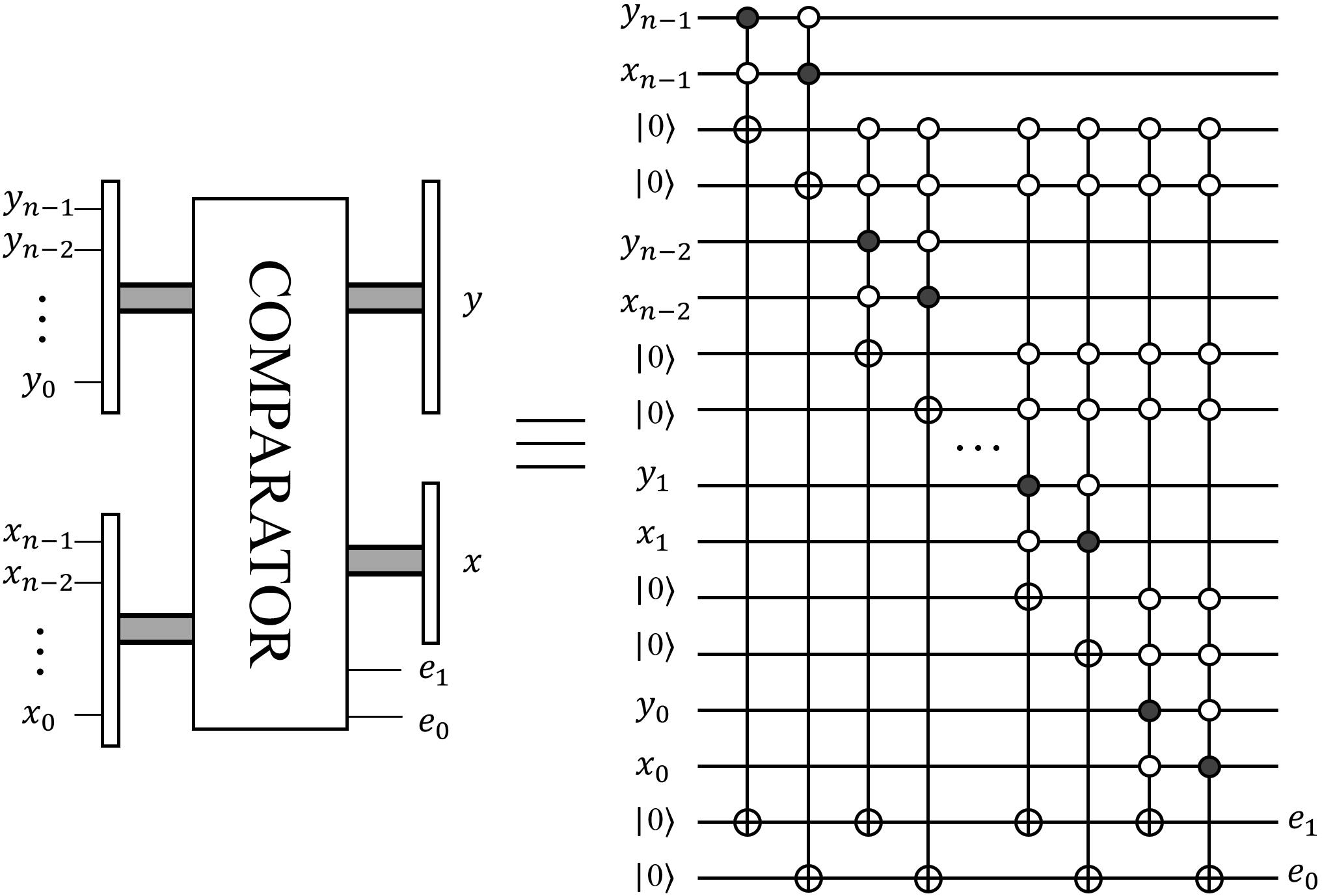}
  \caption{Circuit implementation of quantum COM module (figure and descriptions adapted from \cite{yan2016strategy})}
  \label{fig2}
\end{figure}

 The quantum comparator (i.e. COM) circuit has been widely used in quantum computing literature. Designed in \cite{wang2012design} and as used in \cite{yan2016strategy}, the COM module (Figure \ref{fig2}) compares two states $\vert y\rangle$ and $\vert x\rangle$, where $\vert y\rangle=\vert y_{n-1}\ldots y_1y_0\rangle$ and $\vert x\rangle=\vert x_{n-1}\ldots x_1x_0\rangle$, $y_i,x_i\in\{0,1\}$, $i=0, 1,\ldots, n-1$. Qubits $\vert e_1\rangle$ and $\vert e_0\rangle$ are outputs of the comparison:
\begin{itemize}
  \item If $\vert e_1e_0\rangle=\vert 10\rangle$, then $\vert y\rangle > \vert x\rangle$;
  \item If $\vert e_1e_0\rangle=\vert 01\rangle$, then $\vert y\rangle < \vert x\rangle$;
  \item If $\vert e_1e_0\rangle=\vert 00\rangle$, then $\vert y\rangle = \vert x\rangle$.
\end{itemize}
Therefore, when $\vert e_0\rangle=0$, $\vert y\rangle\geq\vert x\rangle$; otherwise, $\vert y\rangle<\vert x\rangle$. Together with the discussion presented earlier in Subsection \ref{subsec2-2}, we conclude that the SUB module will work on $\vert y\rangle-\vert x\rangle$ only when $\vert e_0\rangle=0$ (i.e. $\vert y\rangle\geq\vert x\rangle$).

\subsection{Quantum accumulator}
\label{subsec2-3}
Since the ghost imaging algorithm requires a series of pixel accumulation operations, we envision the need for circuit network to accumulate these pixels. Hence in this subsection, we present the rudiments for implementing the quantum accumulator (or simply ACC module). Hopefully, the proposed ACC module will be useful for other protocols and applications in the quantum computing domain in general. Mathematically, the ACC module is designed to accomplish the following transformation:
\begin{equation}\label{eq2-x}
\text{ACC} \vert 0\rangle \vert c_{yx}\rangle \vert y\rangle \vert x\rangle =  \Big[\sum_{y=0}^{2^n-1}\sum_{x=0}^{2^n-1}\vert c_{yx}\rangle\Big]\vert c_{yx}\rangle\vert y\rangle\vert x\rangle.
\end{equation}

\begin{figure}[!htb]
  \centering
  \includegraphics[width=12 cm]{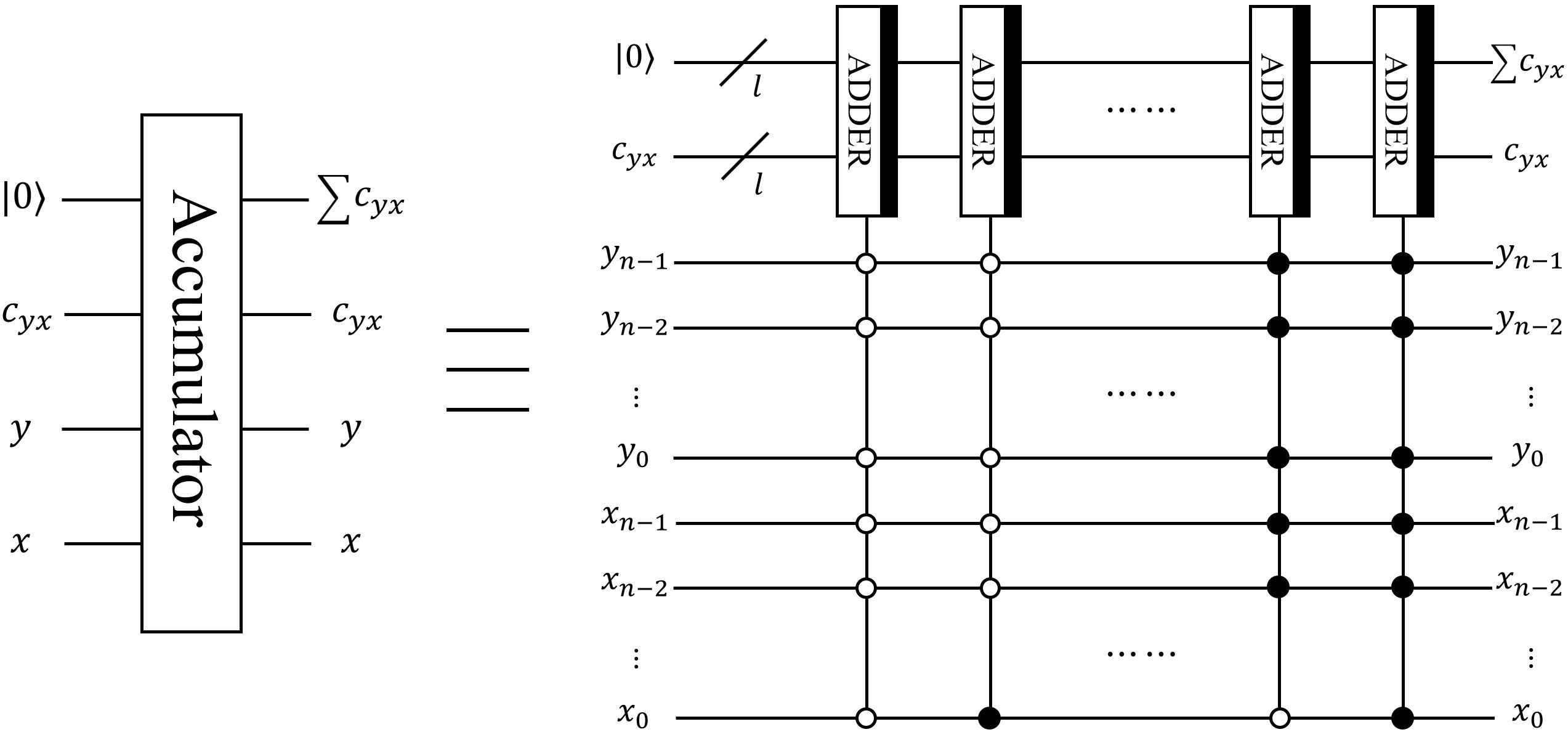}
  \caption{Circuit implementation of quantum ACC module}
  \label{fig3}
\end{figure}

As presented in Figure \ref{fig3}, $\vert y\rangle$ and $\vert x\rangle$, i.e. $\vert y_{n-1}y_{n-2}\ldots y_0\rangle$ and $\vert x_{n-1}x_{n-2}\ldots x_0\rangle$, are 2\emph{n} control qubits of the ADD modules, while $\vert c_{yx}\rangle$ (which includes \emph{l} qubits) stays different state in each pair of $\vert y_ix_i\rangle$. The $\vert y_ix_i\rangle$ ranges from $\vert 0^{\otimes 2n}\rangle$ to $\vert 1^{\otimes 2n}\rangle$, i.e. from $\vert 0\rangle$ to $\vert 2^{2n}-1\rangle$. The $2^{2n}$ ADD modules are utilized to perform the summation of $\vert c_{yx}\rangle$ in each state of $\vert y_ix_i\rangle$. The additional \emph{l} qubits (which are initialized as $\vert 0^{\otimes l}\rangle$) are integrated into the circuit to record the accumulation result of each ADD module, and the combination of these results will be the final output of the whole procedure.

\subsection{Quantum multiplier}
\label{subsec2-4}
Quantum multiplier (i.e. MUL module) is primarily targeted at executing the multiplication operation between two quantum states. Some often-used MUL operations include those in \cite{Vedral1996Quantum} and \cite{yan2017quantum}. In this subsection, although formulated for widespread use, it is designed to support efficient implementation of our proposed QGI protocol. As a premise, we start with the classical multiplication of two binary numbers $y=y_{m-1}\ldots y_1y_0$ and $x=x_{n-1}\ldots x_1x_0$. The calculation process is outlined in Eq. (\ref{eq3}):
\begin{eqnarray}\label{eq3}
yx&=&(y_{m-1}\ldots y_1y_0)x_0 + (y_{m-1}\ldots y_1y_00)x_1 \nonumber \\
&&+\ldots + (y_{m-1}\ldots y_1y_0\overbrace{0\ldots0}^{n-1})x_{n-1} \nonumber \\
&=&yx_0 + (y0)x_1 +\ldots + (y\overbrace{0\ldots0}^{n-1})x_{n-1}
\end{eqnarray}.

To elucidate, let $y=10101$ (i.e. \emph{n}=5) and $x = 1011$ (i.e. $x_3 = 1, x_2 = 0, x_1 = 1, x_0 = 1$, and \emph{m}=4), then $ yx = (10101) \times
(1011)= 11100111$. The execution of this operation can be better comprehended via the stepwise implementation in Eq. (\ref{eq4}).
\begin{eqnarray}\label{eq4}
10101\times1011&=&\sum_{i=0}^3 (2^i y)(x_i) \nonumber \\
&=&(2^0 y)(x_0)+(2^1 y)(x_1)+(2^2 y)(x_2)+(2^3 y)(x_3) \nonumber \\
&=&(\overbrace{10101}^y\overbrace{\cdot}^{m-4=0})\overbrace{(1)}^{x_0}+(\overbrace{10101}^y\overbrace{0}^{m-3=1})\overbrace{(1)}^{x_1} \nonumber\\
&&+(\overbrace{10101}^y\overbrace{00}^{m-2=2})\overbrace{(0)}^{x_2}+(\overbrace{10101}^y\overbrace{000}^{m-1=3})\overbrace{(1)}^{x_3} \nonumber\\
&=&10101 + 101010 + 10101000 = 11100111.
\end{eqnarray}

Employing the ADD module (presented in Subsection \ref{subsec2-2}) and Eq. (\ref{eq3}), our proposed MUL operation is executed using Eq. (\ref{eq5}):
\begin{equation}\label{eq5}
\text{MUL} \vert 0\rangle\vert z\rangle\vert y\rangle\vert x\rangle =  \vert yx\rangle \vert y\overbrace{0\ldots 0}^{m-1}\rangle \vert y\rangle\vert x\rangle,
\end{equation}
where a sequence of $\vert 0\rangle$ is used as input to record the product result of $\vert y\rangle$ and $\vert x\rangle$, while $\vert z\rangle$ (including \textit{m}+\emph{n}-1 qubits) is used to store the temporary results in the multiplication process. The circuit for implementing MUL module is presented in Figure \ref{fig4}, which is a concatenation of \textit{m} ADD operators. The procedure of the MUL circuit that computes the result of two binary number multiplication is shown as follows:

(\romannumeral1) Input: Besides the input states of $\vert y\rangle$ and $\vert x\rangle$, additional qubits of $\vert z\rangle$ and $\vert p\rangle$ (which include \emph{m}+\emph{n}-1 and \emph{n} qubits, respectively) are initialized as a sequence of $\vert 0\rangle$, wherein, $\vert p\rangle$ is used to dynamically store the intermediate result of the multiplication in each step, while $\vert z\rangle$ is used to prepare inputs of the ADD module as explained in the sequel.

(\romannumeral2) Iterative addition: During the (\emph{i}+1)th step, \emph{n} Toffoli gates \big[controlled by the $\vert y\rangle$ and $\vert x_i\rangle, (i = 0,1,\dots m-1)$\big] are applied on the state of $\vert z\rangle$ to obtain a result $\vert z\rangle=\vert y0^{\otimes i}x_i\rangle$ which is regarded as an input for the ADD module (i.e. $\text{ADD}_{i+1}$) in this step, while the other input of $\text{ADD}_{i+1}$ comes from the addition result of $\text{ADD}_{i}$ (it is initialized as $\vert 0^{\otimes n}\rangle$ when \emph{i}=0).

For instance, in Step 1 (i.e. \emph{i}=0), if $\vert x_0\rangle=\vert 1\rangle$, we set $\vert z_{n-1}\ldots z_1z_0\rangle=\vert y_{n-1}\ldots y_1y_0\rangle$, so $\vert z\rangle=\vert y\rangle$ in this case. Then, $\vert z\rangle$ and $\vert p\rangle$ (\emph{p}=0) are considered as two inputs of the $\text{ADD}_1$. Following the addition operation, \emph{n} Toffoli gates are employed to reset $\vert z\rangle$ to its original states, i.e. $\vert 0^{\otimes m+n-1}\rangle$.

Step 3: Output: An iterative approach is used to compute the \emph{p}roduct between of $\vert y\rangle$ and $\vert x\rangle$ (including \emph{m}+\emph{n} qubits) such that $\vert p\rangle=\vert p_{(m+n-1)}\ldots p_1p_0\rangle=\vert y\rangle\vert x_0\rangle+\vert y0\rangle\vert x_1\rangle+\ldots +\vert y0^{\otimes{m-2}}\rangle\vert x_{m-2}\rangle+\vert y0^{\otimes{m-1}}\rangle\vert x_{m-1}\rangle$.


\addfigure{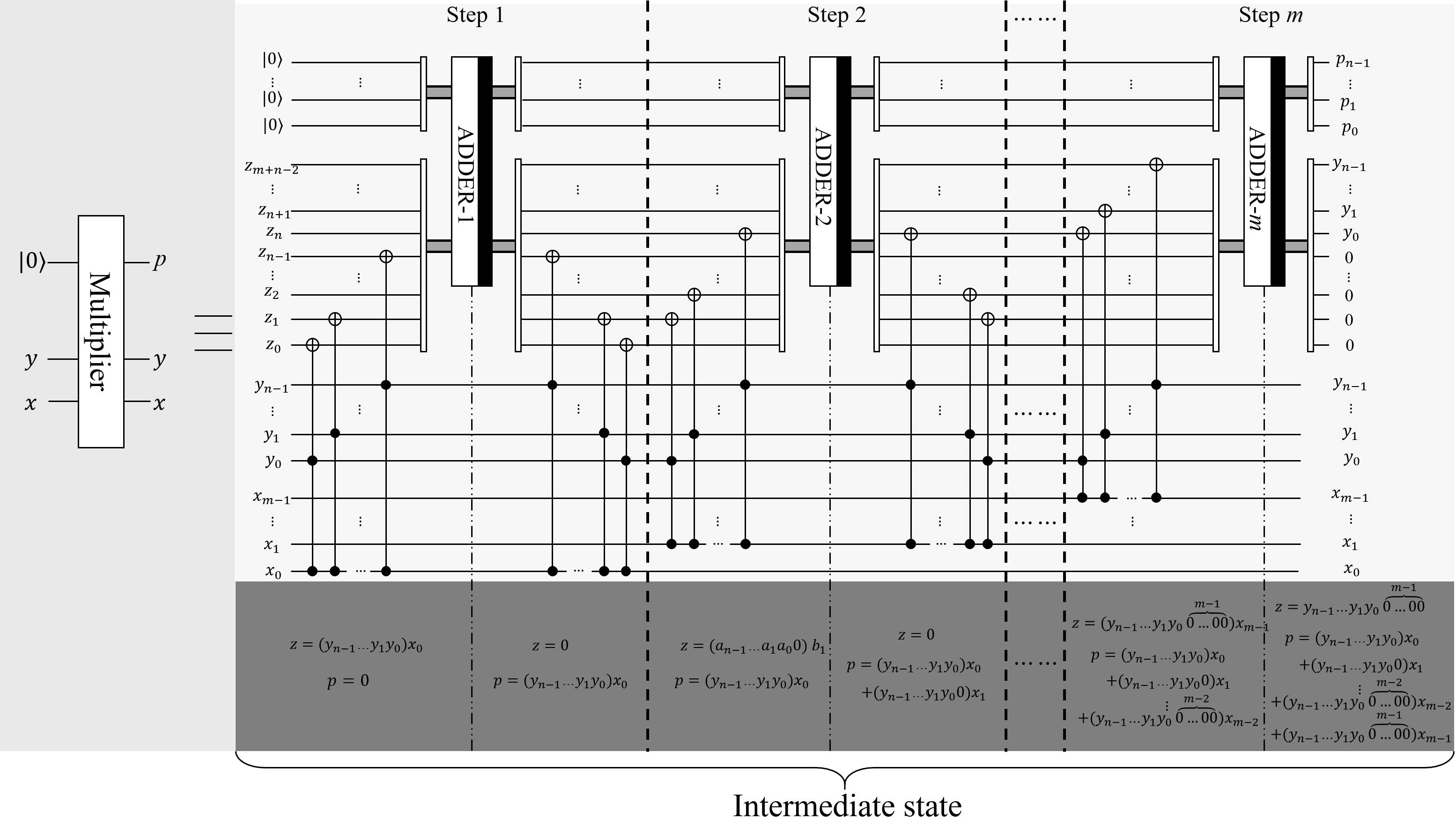}{Circuit implementation of quantum MUL module}{fig4}{width=1.0\textwidth}

\subsection{Quantum divider}
\label{subsec2-5}
On classical computers, the binary division operation is actually a series of subtraction tasks. Consider, as an example, the classical division operation ``$\frac{100110}{110}$" (as presented in Figure \ref{fig5}), which can be executed via the following 4 steps:

Step 1: A set of the first three bits in the numerator (i.e. the dividend) ``100" is taken as minuend in this step, which is used to compare with the denominator (i.e. the divisor) ``110" using the COM module (as presented in Subsection 2.3). Since $100<110$, the subtraction operation is not applicable. Therefore, the result of the Step 1 becomes ``100'' with the next bit in the numerator (in this case ``1'') making a sequence ``1001'' as the new minuend sequence.

Step 2: The two binary sequences being compared, in this step, are the outcome from Step 1 (i.e. ``1001") and the denominator (i.e. ``110"). Since $110<1001$, the SUB module (as presented in Section 2.2) will be applied to perform the subtraction. The result of this subtraction ``11'' and the next bit in the numerator sequence (i.e. ``1") would then serve as the minuend (i.e. ``111'') in the next step of the operation.

Step 3: Like in the previous steps, we compare two states using the COM module. However, in this case, the minuend resulting from Step 2 (i.e. ``111") is compared with the divisor (i.e. ``110"). Since $111>110$, we proceed with the subtraction $111-110=01$. Finally, the last bit of the numerator (i.e. ``0") is juxtaposed with the outcome of this subtraction to form ``010'', which will serve as the minuend for the next step of the operation.

Step 4: Similarly, we compare the minuend ``010'' with the divisor ``110'', and since $010<110$, the subtraction operation is not activated. Having exhausted the bits in the numerator sequence, our operation returns the last two bits of the minuend ``10" as the reminder of the division operation. In the event of the opposite scenario, i.e. the minuend is greater than or equal to ``110", the SUB operation is used to obtain the difference that serves rest of the division operation.

As outlined in the four steps above, the division ``$\frac{100110}{110}$" produces a quotient ``110" and a remainder of ``10". Figure \ref{fig6} presents a pictorial implementation of the four steps outlined earlier. Based on them, we propose a quantum divider (i.e. DIV module) to implement the division operation.

\addfigure{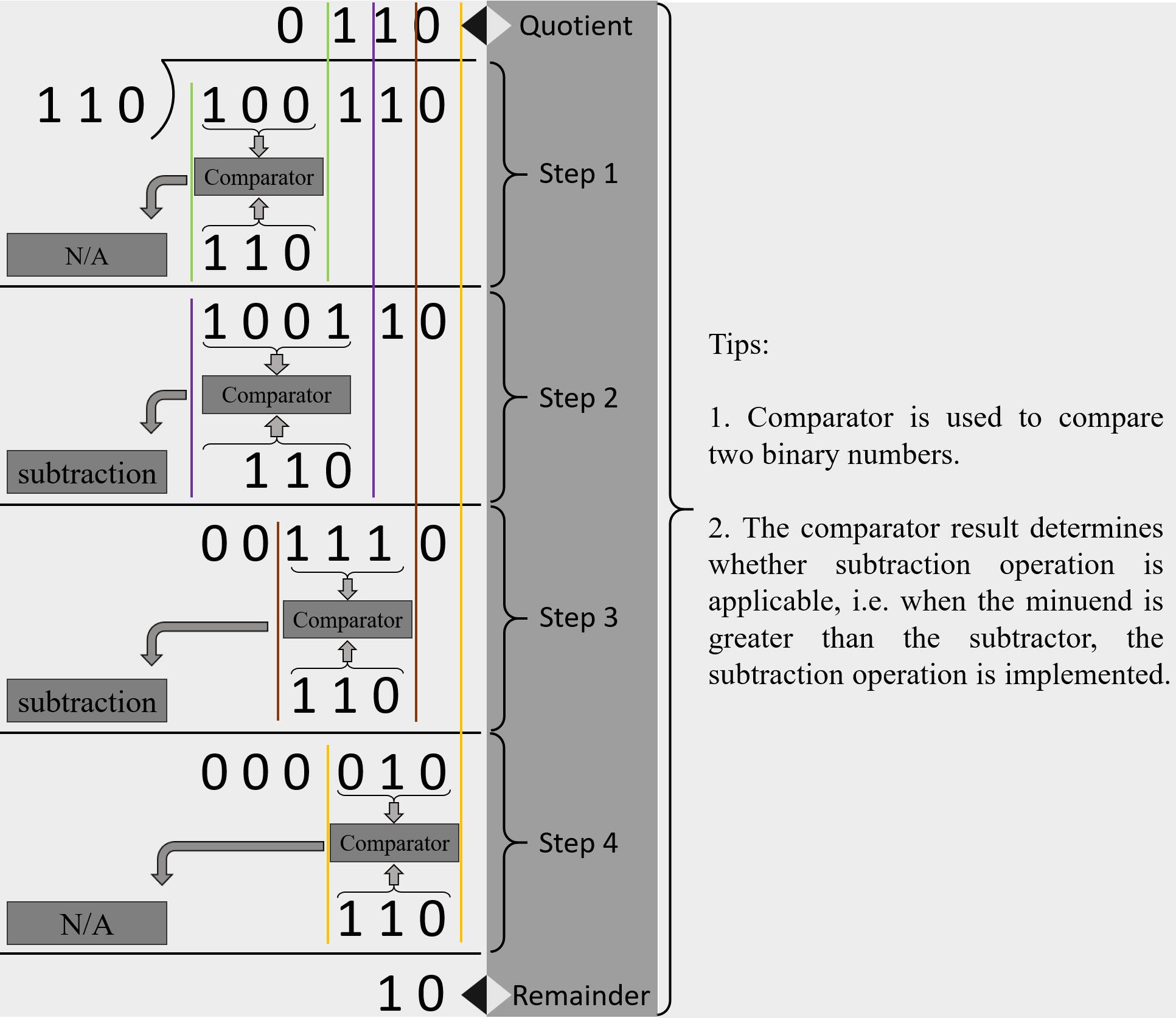}{Illustration of the binary division operation}{fig5}{width=0.6\textwidth}

An overview of the composition, formulation, and circuitry to implement the quantum DIV module that executes the division operation is presented forthwith. Consider a sequence $\vert y\rangle=\vert y_{m-1}\ldots y_1y_0\rangle$ as the dividend (numerator) and another one $\vert x\rangle=\vert x_{n-1}\ldots x_1x_0\rangle$ as the divisor (denominator) of a division operation. Then, using a depository, additional information $\vert x^\prime\rangle=\vert x^\prime_{m-1}\ldots x^\prime_1x^\prime_0\rangle$ emanating from the stepwise execution of the subtraction operation (itself part of the quantum DIV operation) to divide $\vert y\rangle$ by $\vert x\rangle$, the result of which (i.e. \textit{q}uotient) is recorded as $\vert q\rangle=\vert q_{m-n+1}\ldots q_1q_0\rangle$.

As presented in Step 1 of the DIV circuit (in Figure \ref{fig6}), the first \emph{n} CNOT gates are used to map the state of $\vert x_{n-1}\ldots x_1x_0\rangle$ to $\vert x^\prime_{m-1}x^\prime_{m-2}\ldots x^\prime_{m-n}\rangle$. Subsequently, the COM module (presented earlier in Subsection \ref{subsec2-c}) is utilized to compare between states $\vert y_{m-1}\ldots y_1y_0\rangle$ and $\vert x^\prime_{m-1}x^\prime_{m-2}\ldots x^\prime_0\rangle$. An useful state comes out the COM module is $\vert e_0\rangle$. A result $\vert e_0\rangle=\vert 0\rangle$ (i.e. $\vert y\rangle \geq \vert x^\prime\rangle$) activates the $e_0$-controlled SUB module to obtain the subtraction outcome of $\vert v_1\rangle=\vert y\rangle$-$\vert x^\prime\rangle$ (as the \emph{input} of COM module in Step 2). Otherwise, the SUB module is inactive so the outcome remains $\vert y\rangle$ (See the full illustration at the bottom of Figure \ref{fig6}). Following the SUB module, the 1st $e_0$-controlled CNOT gate is applied on $\vert q_{m-n}\rangle$ to obtain the first (also leftmost) or most significant qubit of the division result (i.e. quotient), while the 2nd CNOT gate ensures $\vert e_0\rangle=\vert 0\rangle$ before entering Step 2. At the end of the Step 1, additional \emph{n} CNOT gates are similarly used to reset state $\vert x^\prime\rangle$ to its initialized state, i.e. sequence of $\vert 0^{\otimes m}\rangle$ entries, preparatory for its use in Step 2.

In Step 2, the first \emph{n} CNOT gates assign the value $\vert 0x0^{\otimes{m-n-1}}\rangle$ to qubit $\vert x^\prime\rangle$ for the comparison with the \emph{output} from Step 1 (i.e. $\vert v_1\rangle$). Similarly, $\vert e_0\rangle=\vert 0\rangle$ indicates $\vert v_1\rangle\geq \vert x^\prime\rangle$ so that the subtraction operation will be executed producing an outcome $\vert v_2\rangle=\vert v_1\rangle-\vert x^\prime\rangle$. Following that, the next (\emph{n}+2) CNOT gates are applied to (1) obtain the 2nd qubit of the division result, (2) assure $\vert e_0\rangle=\vert 0\rangle$ in the next step, and (3) reset $\vert x^\prime\rangle$ from $\vert 0x0^{\otimes{m-n-1}}\rangle$ to $\vert0^{\otimes m}\rangle$. The final outcome of the DIV module (indicated as $\vert q\rangle$ at the end of the circuit) is the iterative execution of the steps enumerated above. Meanwhile, the sequence $\vert v_{m-n+1}\rangle=\vert v_{m-n}\rangle-\vert x^\prime\rangle$ (technically, composed of $\vert r_{n-1}r_{n-2}\ldots r_0\rangle$ at the end of the circuit) is regarded as the remainder from the DIV operation.

\addfigure{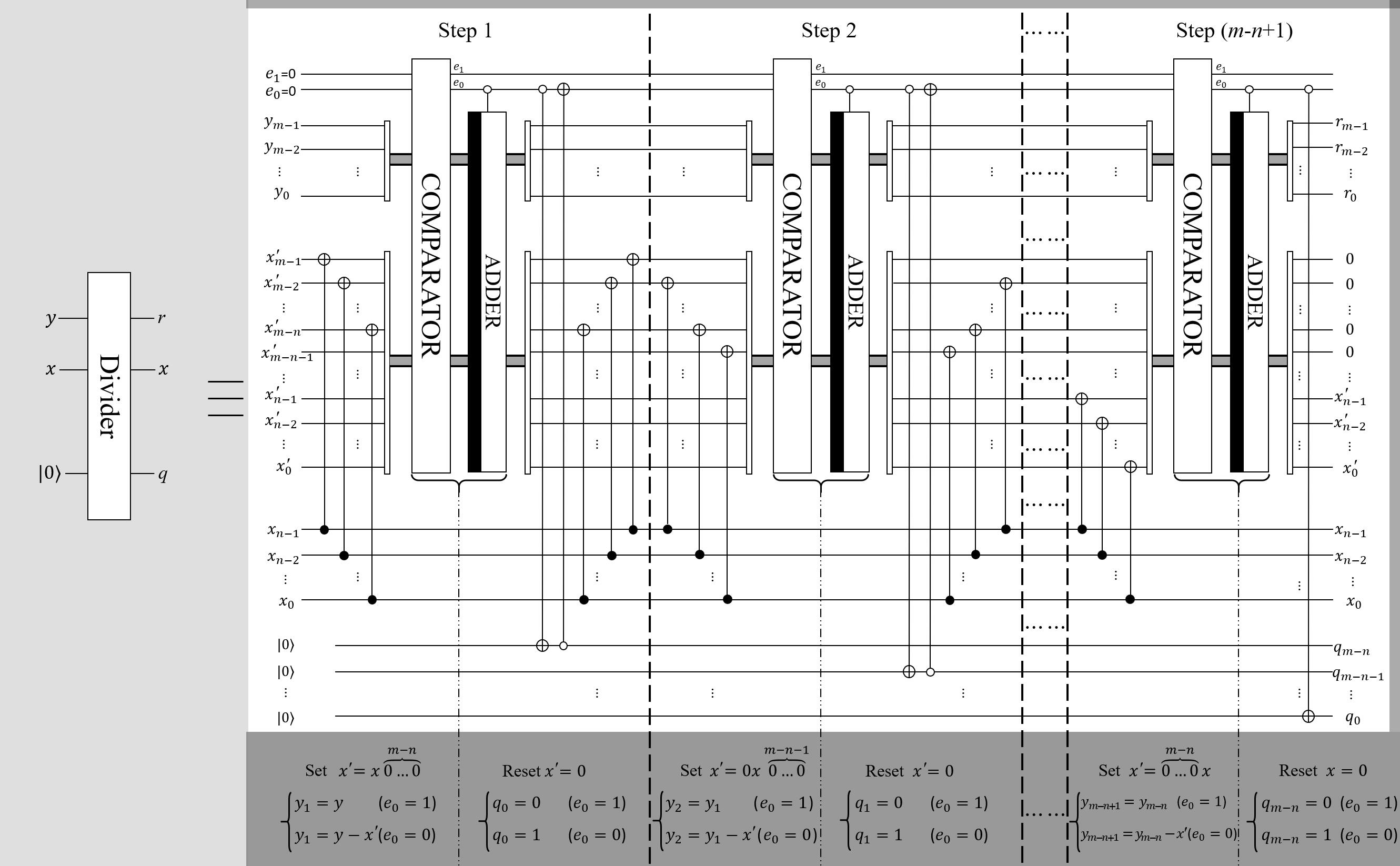}{Circuit implementation of quantum DIV module}{fig6}{width=1.0\textwidth}

\section{Circuit-model implementation of ghost imaging experiment}
\label{sec3}
\subsection{Mathematical formulation of ghost imaging}
\label{subsec3-1}
Following the mathematical discussions in \cite{Basano2007A}, the methodical outline of our proposed quantum ghost imaging experiment is described in Figure \ref{fig7}. There are two speckle pattern sequences, i.e. $I_{2^s-1}^gI_{2^s-2}^g\ldots I_k^g\ldots I_1^gI_0^g$ and $I_{2^s-1}^dI_{2^s-2}^d\ldots I_k^d\ldots I_1^dI_0^d$, wherein, $I_k^g$ and $I_k^d$ indicate the speckle patterns in the si\emph{g}nal and i\emph{d}ler fields, respectively. The speckle patterns at the same position of two sequences are identical and each pattern consists of $2^m\times 2^n$ pixels in the region of $\mathbb{U}$. $\mathbb{U}_k(y,x)$ represents a pixel whose coordinate is (\emph{y},\emph{x}) at the \emph{k}th sample pattern (i.e. $I_k^g$ in the signal field) in the sequence, where $y\in\{0,1,\ldots, 2^m-1\}$, $x\in \{0,1,\ldots, 2^n-1\}$, and $k\in\{0,1, \ldots, 2^s-1\}$. We define a subregion $\mathbb{A}$ of $\mathbb{U}$, where $\mathbb{A}\subseteq\mathbb{U}$, i.e. the pixels in $\mathbb{A}$ are subset of pixels in $\mathbb{U}$. Furthermore, we define a characteristic function $\mathcal{F}_k(y,x)$ in the region $\mathbb{U}$, if $\mathbb{U}_k(y,x)$ is in the sub-region $\mathbb{A}$, then $\mathcal{F}_k(y,x)=1$; otherwise $\mathcal{F}_k(y,x)=0$.

\addfigure{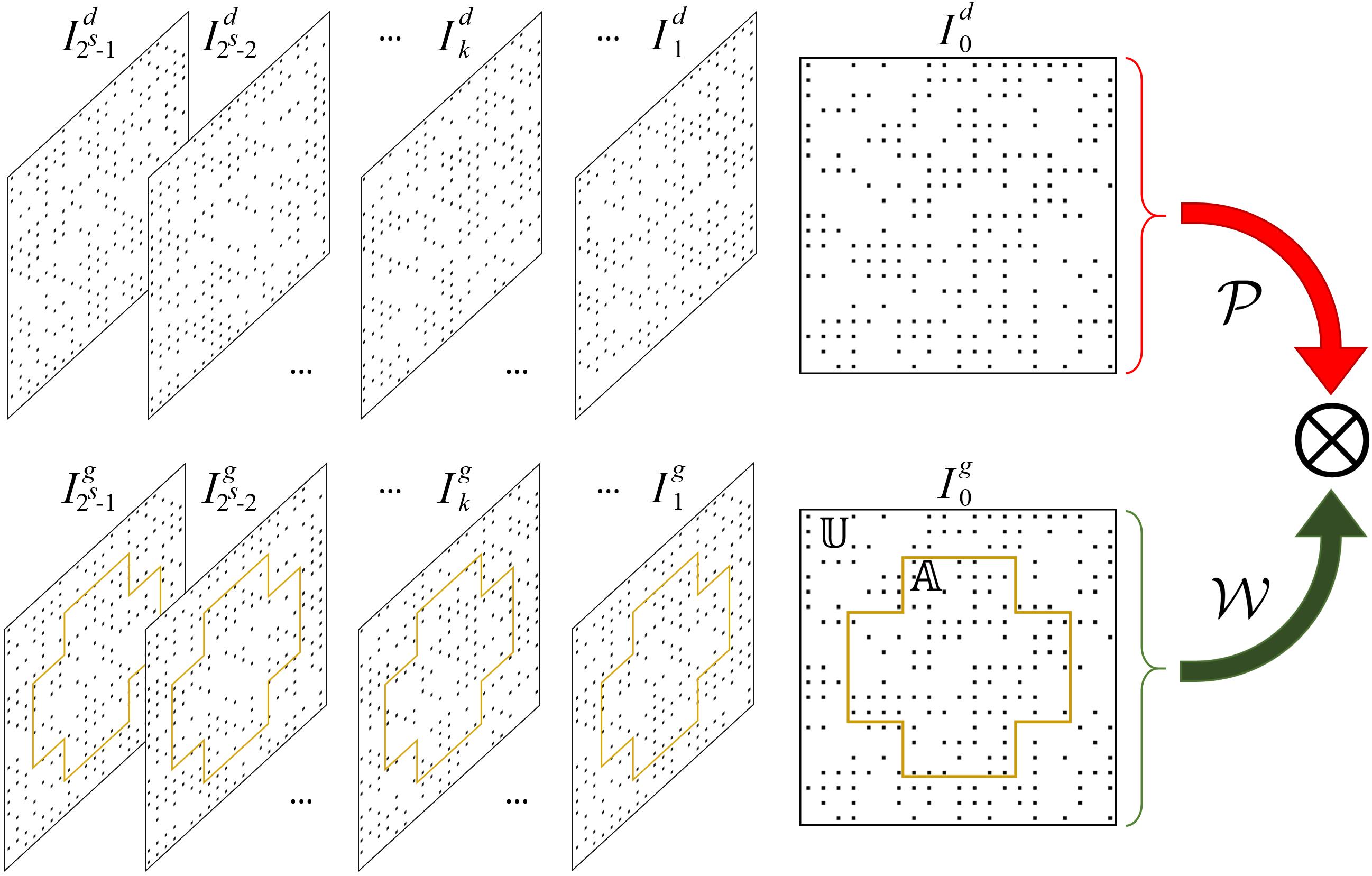}{Mathematical analysis of the setups of ghost imaging experiment}{fig7}{width=0.8\textwidth}

We define a special vector $\mathcal{W}(k)$ whose \emph{k}th element records the statistical weight for the \emph{k}th speckle pattern in the sequence (in the signal field). Formally, $\mathcal{W}(k)$ could be calculated as:
\begin{equation}\label{eq6}
\mathcal{W}(k)=\sum_{y=0}^{2^m-1}\sum_{x=0}^{2^n-1}\mathcal{F}_k(y,x)\mathbb{U}_k(y,x),\ \ \ \ \ (k=0,1,\ldots,2^s-1).
\end{equation}
where $\mathcal{W}(k)$ calculates the sum of pixels belonging to $\mathbb{A}$ in \emph{k}th speckle pattern. It represents the weight of the \emph{k}th speckle pattern in the signal field in the ghost imaging experiment. While $\mathcal{P}$ (in Figure \ref{fig7}) indicates the sequence of speckle patterns that are obtained from the idler field. ``$\bigotimes$" in Figure \ref{fig7} performs cross-correlation operation between the speckle patterns $\mathcal{P}$ and their statistical weights $\mathcal{W}$, after which, the ghost imaging result $\mathcal{R}$ is obtained in the form formulated in Eq. (\ref{eq7}):
\begin{equation}\label{eq7}
\mathcal{R}=\langle\mathcal{W}\mathcal{P}\rangle-\langle\mathcal{W}\rangle\langle\mathcal{P}\rangle,
\end{equation}
where $\langle\cdot\rangle=\frac{1}{2^s-1}\sum_k\cdot$ denotes an ensemble average over $2^s-1$ phase realizations. Meanwhile, every pixel in $\mathcal{R}$ (e.g., at coordinate $(y,x)$) could be represented as:
\begin{equation}\label{eq8}
\mathcal{R}(y,x)=\frac{1}{2^s-1}\sum_{k=0}^{2^s-1}\mathcal{W}_k\mathcal{P}_k(y,x)-\frac{1}{2^s-1}\sum_{k=0}^{2^s-1}\mathcal{W}_k\cdot\frac{1}{2^s-1}\sum_{k=0}^{2^s-1}\mathcal{P}_k(y,x).
\end{equation}

The remainder of this section dwells on the quantum circuit implementation of ghost imaging based on the formulations and discussions presented in Eq. (\ref{eq6})-(\ref{eq8}) as well as earlier sections of the study.

\subsection{Establishment of quantum speckle patterns}
\label{subsec3-2}
A speckle pattern is defined as a quantum image as presented in Eq. (\ref{eq1}) (where \emph{l}=1), i.e. the chromatic information of every pixel in the speckle pattern only includes two levels, i.e. 0 (black) or 1 (white). It is randomly generated by utilizing the Hadamard gate, i.e.
$H=\frac{1}{\sqrt{2}}
\small
\begin{pmatrix}
1&&1\\-1&&1
\end{pmatrix}$,
that could transform the initial state $\vert 0\rangle$ to $\frac{1}{\sqrt{2}}(\vert 0\rangle + \vert 1\rangle)$. The whole procedure is outlined in the circuit in Figure \ref{fig8}. It is noteworthy that, whereas on classical computers, the randomness of the gray value is generated with a certain periodicity (i.e. it is pseudo-random), by utilizing the quantum gates (Hadamard and CNOT gates), real-random of speckle patterns can be generated on the quantum computing paradigm.

Moreover, as presented in Figure \ref{fig8}, all $\vert 0\rangle$ initial states are transferred into the outcome of the speckle pattern, which consists of $\vert s\rangle=\vert s_{t-1}s_{t-2}\ldots s_0\rangle=\vert 0\rangle^{\otimes t}$, $\vert y\rangle=\vert y_{m-1}y_{m-2}\ldots y_0\rangle=\vert 0\rangle^{\otimes m}$, and $\vert x\rangle=\vert x_{n-1}x_{n-2}\ldots x_0\rangle=\vert 0\rangle^{\otimes n}$ (all together $m+n+t$ qubits), while an additional qubit $\vert c_{yx}\rangle$ is employed to record the gray value of every pixel in these patterns. The transformation starts by applying a cortege of Hadamard gates on every qubit of $\vert 0\rangle^{\otimes m+n+t}$, i.e. $\mathcal{H}=H^{\otimes m}\otimes H^{\otimes n}\otimes H^{\otimes t}$ on $\vert 0\rangle^{\otimes m+n+t}=\vert 0\rangle^{\otimes m}\otimes \vert 0\rangle^{\otimes n}\otimes \vert 0\rangle^{\otimes t}$, as formulated in Eq. (\ref{eq10}):
\begin{eqnarray}\label{eq10}
\mathcal{H}(\vert 0\rangle^{\otimes m+n+t})&=&\frac{1}{2^{(m+n+t)/2}}(\overbrace{\vert 0\ldots 00\rangle}^{(m+n+t)\text{qubits}}+\overbrace{\vert 0\ldots 01\rangle}^{(m+n+t) \text{qubits}}+\ldots+\overbrace{\vert 1\ldots11\rangle}^{(m+n+t) \text{qubits}}) \nonumber \\
&=& \frac{1}{2^{(m+n+t)/2}}(\sum_{y=0}^{2^m-1}\vert y\rangle\otimes \sum_{x=0}^{2^n-1}\vert x\rangle\otimes \sum_{s=0}^{2^t-1}\vert s\rangle).
\end{eqnarray}

After the processing in Eq. (\ref{eq10}), the initial state has been transformed to an intermediate state, which represents a sequence of speckle patterns ($2^t$ patterns) where each pattern is a $2^m\times 2^n$ quantum image. Finally, a Hadamard gate is utilized on the chromatic wire to assign a binary value to every pixel in the series of speckle patterns. Upon executing Eq. (\ref{eq10}), the initial state has been transformed to an intermediate state that evolves into the final speckle pattern composed of states $\vert y\rangle$, $\vert x\rangle$, $\vert s\rangle$, and $\vert c_{yx}\rangle$ as shown in Figure \ref{fig8}.

\addfigure{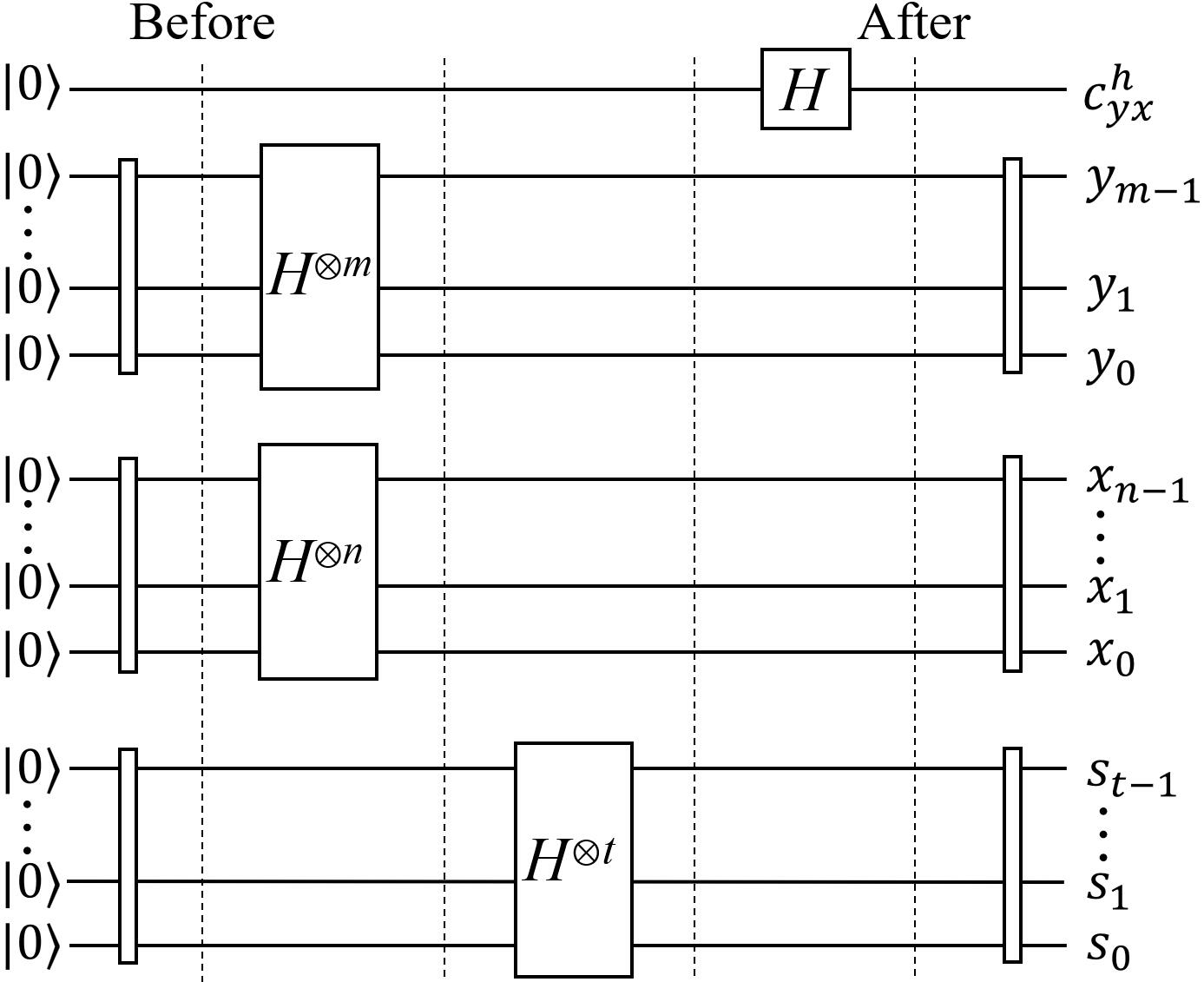}{Circuit realization of the quantum speckle patterns}{fig8}{width=0.6\textwidth}

\subsection{Quantum circuit for ghost imaging experiment}
\label{subsec3-3}
Ghost imaging is a technique that is focused on producing an image of an object by combining information from two other sources. In photonic quantum computing, implementation of ghost imaging involves the use of source of pairs of entangled photons and each pair is shared between the two detectors. Since the outcome of Section 3.1, our QIP-based circuit model for implementing QGI could be construed by referring to Eq. (\ref{eq7}). Consequently, we tailor our QGI in terms of using quantum computing resources to compute the parameters ($\langle\mathcal{W}\mathcal{P}\rangle$, $\langle\mathcal{W}\rangle$, $\langle\mathcal{P}\rangle$, and $\langle\mathcal{W}\mathcal{P}\rangle-\langle\mathcal{W}\rangle\langle\mathcal{P}\rangle$) in Eq. (\ref{eq7}).

\subsubsection{Calculation of $\langle\mathcal{W}\mathcal{P}\rangle$ and its circuit implementation}
\label{subsubsec3-3-1}
In quantum information and quantum computing, circuit is an effective and pictorial description to the quantum state evolution from the inputs to its outputs. The calculation of $\langle\mathcal{W}\mathcal{P}\rangle$ is detailed in the remainder of this subsection, we outline the execution of the five stages of the quantum circuit implementation of the QGI as presented in Figure \ref{fig9}.

\begin{description}
\item[(\romannumeral1)]Interaction between the speckle patterns and phase mask
    \\
    As delineated in blue rectangle in Figure \ref{fig9}, this unit specifies the interaction between the speckle patterns and the phase mask so as to obtain the quantum \emph{i}nterested image (i.e. $\vert I_i\rangle$). The qubit sequence $\vert y_{m-1}\ldots y_1y_0\ x_{n-1}\ldots x_1x_0\ c_{yx}^m\rangle$ at the top of this unit represents the quantum phase mask, wherein $\vert y\rangle$ and $\vert x\rangle$ are the coordinates, while $\vert c_{yx}^{m}\rangle$ records the gray value of the pixel in the quantum \emph{m}ask image. The extra qubits, i.e. $\vert y\rangle$, $\vert x\rangle$, $\vert s\rangle$, and $\vert c_{yx}^{s}\rangle$, represent a series of speckle patterns, where $\vert s\rangle$ indicates the number of the patterns, and $\vert y\rangle$ as well as $\vert x\rangle$ is the pixel coordinates of the pattern, while $\vert c_{yx}^s\rangle$ represents the gray value of pixels in these \emph{s}peckle patterns. In addition, $\vert c_{yx}^{i}\rangle$ records the gray value that the patterns across the mask image, i.e. the generated quantum \emph{i}nterested quantum image.

    To realize the needed interaction between the quantum mask image and the quantum speckle pattern, as discussed in Subsection \ref{subsec3-1}, a position-wise correspondence comparison between the pixels at the same position of the quantum mask image and each quantum speckle pattern is required. To do this, we have to judge the gray values of the two pixels (i.e. $\vert c_{yx}^{m}\rangle$ and $\vert c_{yx}^{s}\rangle$), if both of them are $\vert 1\rangle$, the gray value of the pixel in the interested image (i.e. $\vert c_{yx}^{i}\rangle$) is transformed from its initialized state $\vert 0\rangle$ to $\vert 1\rangle$ by using a CNOT gate (with multiple control conditions). Otherwise, the gray value of the pixel in the interested image remains $\vert 0\rangle$ (i.e. black). The control conditions imposed on the $\vert y\rangle$ and $\vert x\rangle$ coordinates of the quantum mask image are identical with those in the quantum speckle pattern to ensure the operation happens at the same position of them.

\item[(\romannumeral2)]Calculation of the weight $\bm{\vert w_s\rangle}$ in the signal field
    \\
    The second unit of our circuit (with yellow legend) accumulated the weights of all the pixels of each quantum interested image. The quantum ACC module presented in Subsection \ref{subsec2-3} is used to undertake this task. The qubit sequence $\vert y\rangle$ and $\vert x\rangle$ as well as $\vert c_{yx}^i\rangle$ are regarded as three inputs of the ACC module, while $\vert w_s\rangle=\sum_{y=0}^{2^m-1}\sum_{x=0}^{2^n-1}\vert c_{yx}^i\rangle$ is the output to store the weight value of every quantum interested image. The control qubits applied on the ACC module in Figure \ref{fig9} assure that the pixel accumulation happens within the same interested image, i.e. the weight $\vert w_s\rangle$. After the accumulation operation, the output of this part is $\vert s^w w_s\rangle$ which contains all the weights of each quantum interested image.

\item[(\romannumeral3)]Calculation of the product of weights and speckle patterns
    \\
    The third unit of our circuit (highlighted in green) is used to compare the product of weights ($\vert w_s\rangle$ in the signal field) and the sequence of speckle patterns in the idler filed ($\vert I_d\rangle$) which are not interacted with the mask. The quantum MUL module presented in Subsection \ref{subsec2-4} is used to perform the calculation. The two inputs of the MUL module are the weights of the signal field $\vert w_s\rangle$ and $\vert c_{yx}^s\rangle$ (i.e. the chromatic information of $\vert I_d\rangle$). Meanwhile, $\vert w_s\rangle$ should be multiplied with all the pixels of a speckle pattern ($\vert I_k^d\rangle$) in the sequence. The control condition operation on $\vert s^w\rangle$ and $\vert s\rangle$ confine the operation in the same pairs in the multiplication process (i.e. each $\vert w_s\rangle$ corresponds to each $\vert I_k^d\rangle$). The outcome is stored in $\vert c_{yx}^s\rangle=\vert c_{yx}^s\rangle\vert w_s\rangle$ and used as the input of the next unit.

\item[(\romannumeral4)]Accumulation of the corresponding pixels in the speckle pattern sequence
    \\
    At the top of the part in purple color, a quantum ACC module is used. The inputs of ACC module are $\vert c_{yx}^s\rangle$ and $\vert s\rangle$, while the output is stored in $\vert c_{yx}^p\rangle$. The purpose of this unit is to simultaneously accumulate scale of all the pixels' gray value in the same position of each pattern (in the sequence). The control condition on $\vert y\rangle$ and $\vert x\rangle$ ensure that the ACC module is restricted to the same position-wise concurrence in each pattern.

\item[(\romannumeral5)]Calculation of the final result $\vert \langle w_sp_s\rangle\rangle$
    \\
    The chromatic information of the speckle patterns in the idler field, i.e. $\vert c_{yx}^{p}\rangle$, has been turned to $\vert w_sp_s\rangle$ after $2^{2n}$ accumulation operation from the last part. The main task of this part is to change the $\vert w_sp_s\rangle$ to $\vert \langle w_sp_s\rangle\rangle$, i.e. the ensemble average over $\vert s\rangle$ phase realizations of $\vert w_sp_s\rangle$. Using the DIV module proposed in Subsection \ref{subsec2-5}, the state $\vert w_sp_s\rangle$ could be divided by $\vert s\rangle$ to obtain the final result, i.e. $\vert\langle w_sp_s\rangle\rangle$.

\end{description}

\addfigure{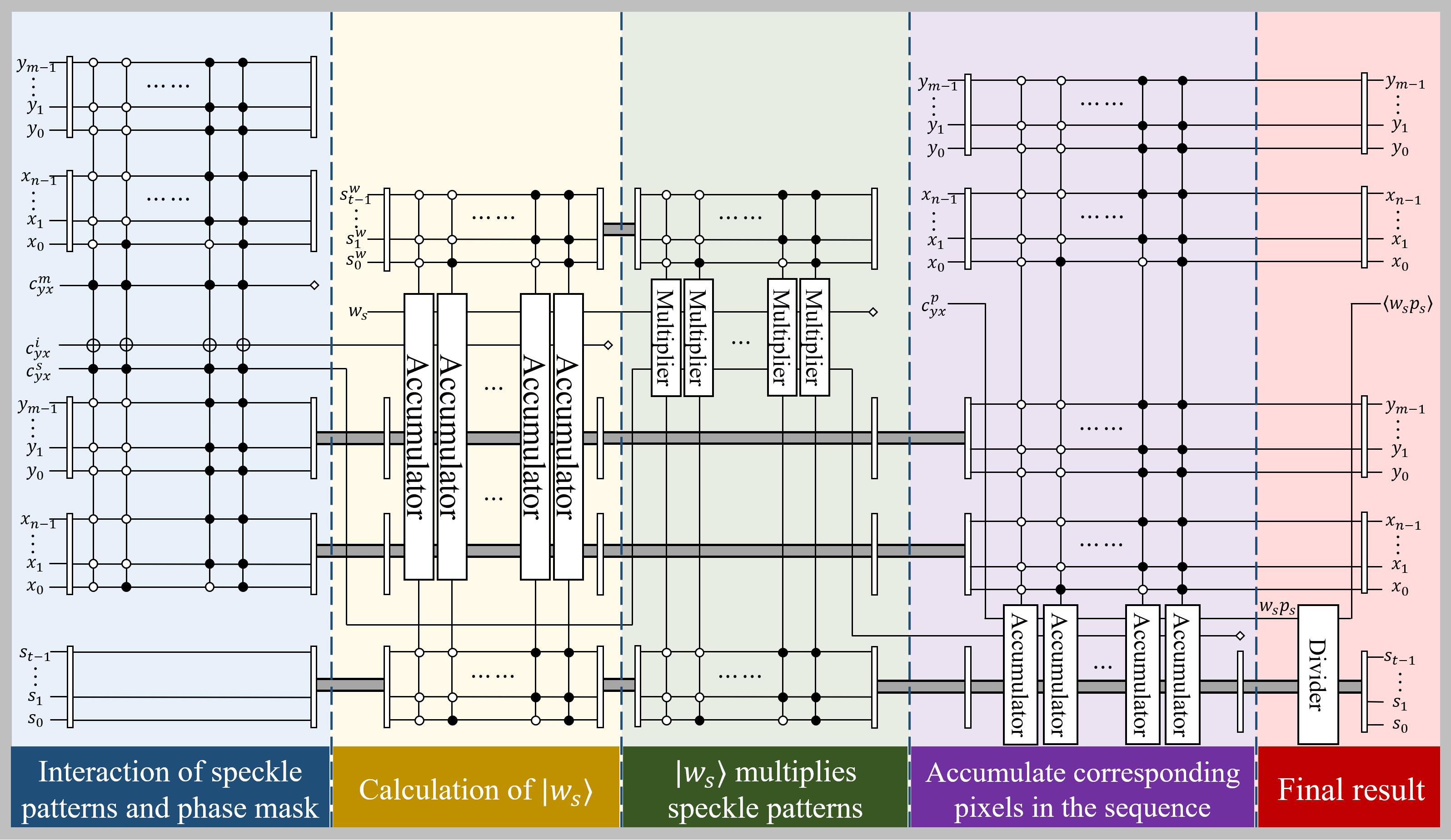}{Circuit realization of $\vert\langle w_sp_s\rangle\rangle$}{fig9}{width=1.0\textwidth}

\subsubsection{Calculation of $\vert\langle\mathcal{P}\rangle\rangle$ and $\vert\langle\mathcal{W}\rangle\rangle$ and their circuit implementations}
\label{subsubsec3-3-2}
The calculation circuit of $\vert \langle p_s\rangle \rangle$ is shown in Figure \ref{fig10}(a). The $\vert yxc_{yx}^p\rangle$ is initialized as $\sum_{y=0}^{2^m-1}\vert y\rangle\otimes \sum_{x=0}^{2^n-1}\vert x\rangle\otimes \vert 0\rangle$ which stores the final result $\vert \langle p_s\rangle\rangle$. In addition, $\vert c_{yx}^s y x s\rangle$ represents the quantum speckle patterns which is created by the circuit as shown in Figure \ref{fig8} and its state is equal to the state $\vert c_{yx}^s y x s\rangle$ as shown in Figure \ref{fig9}. The concurring chromatic information of each pixel ($\vert c_{yx}^s\rangle$) at the same position in $2^s$ speckle patterns are accumulated using ACC module and its result would be stored in the state of $\vert p_s\rangle$. Meanwhile, $2^{2n}$ control qubits combinations guarantee concordance in terms of the content of these patterns. In addition, $\vert p_s\rangle$ is the input of DIV module to calculate the final result, i.e. $\vert \langle p_s\rangle\rangle=\vert p_s\rangle/\vert s\rangle$.

Figure \ref{fig10}(b) outlines the calculation circuit of $\vert\langle w_s\rangle\rangle$ as well as its discussions. The interaction between the speckle patterns and the quantum mask image, and the calculation of $\vert w_s\rangle$ utilizing $2^m$ ACC modules own the same process procedures with Figure \ref{fig9} in blue and yellow color. In addition, $\vert w_s\rangle$ is a superposition state of $2^m$ states corresponding to each state of $\vert s^w\rangle$. The ACC module in the last unit (highlighted in red rectangle) is used to aggregate all the $\vert w_s\rangle$ states. Finally, The accumulation result is divided by $\vert s\rangle$ to calculate the final result, i.e. $\vert \langle w_s\rangle \rangle$.

\addfigure{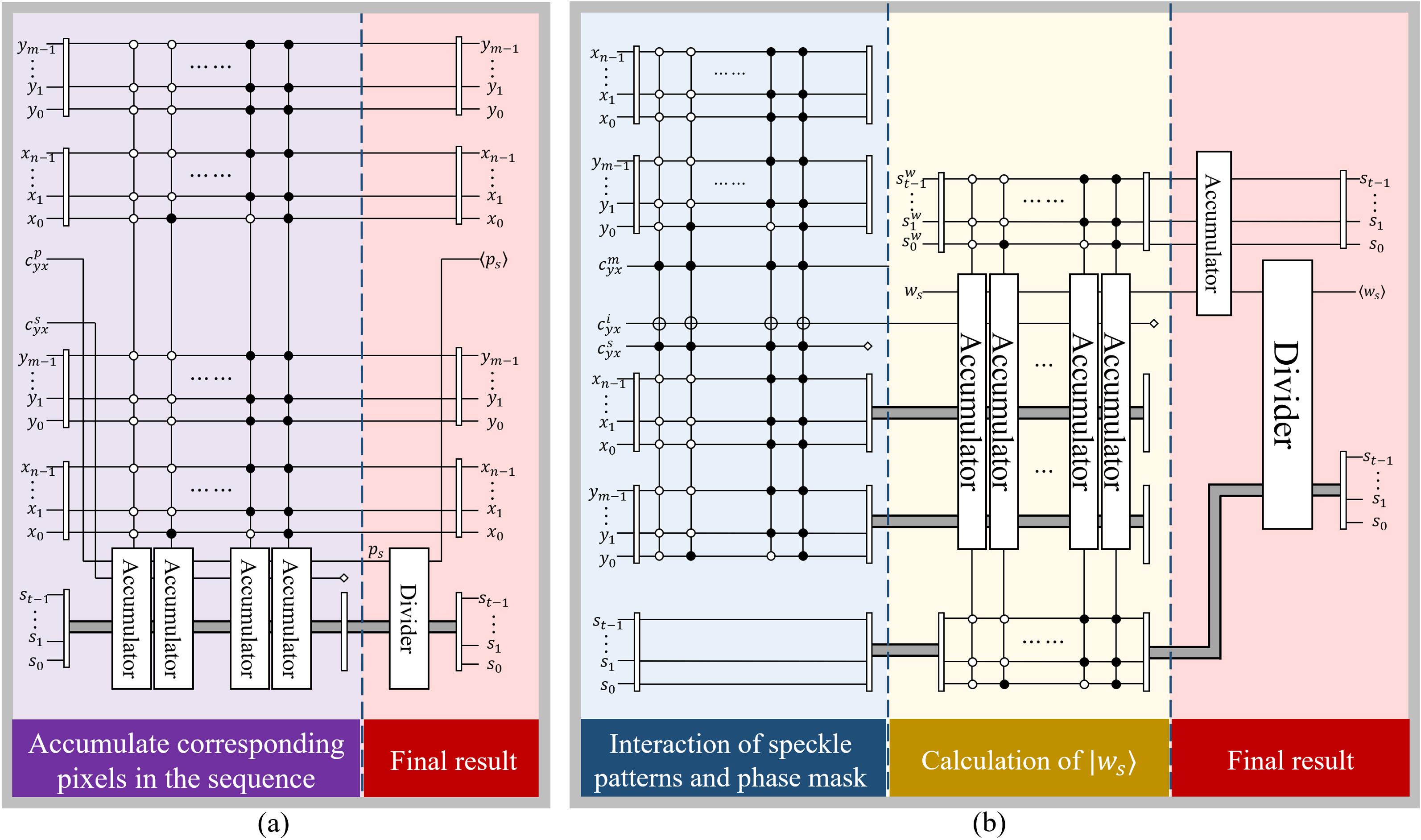}{Circuit realization of (a) $\vert \langle p_s\rangle\rangle$ and (b) $\vert \langle w_s\rangle\rangle$}{fig10}{width=1.0\textwidth}

\subsubsection{Creation of the ghost image by using $\vert \mathcal{R}\rangle=\vert\langle\mathcal{W}\mathcal{P}\rangle\rangle-\vert\langle \mathcal{W}\rangle\rangle\vert\langle\mathcal{P}\rangle\rangle$ }
\label{subsubsec3-3-3}
By concatenating the various sub-circuits presented in this section, we realize the circuit to implement the ghost imaging technique which also translates to the operation $\vert \mathcal{R}\rangle = \vert\langle\mathcal{W}\mathcal{P}\rangle\rangle- \vert\langle\mathcal{W}\rangle\rangle\vert\langle\mathcal{P}\rangle\rangle$ (that is $\vert\langle r_s\rangle\rangle=\vert\langle w_sp_s\rangle\rangle-\vert\langle w_s\rangle\rangle\vert\langle p_s\rangle\rangle$ as presented in Figure \ref{fig11}). It is trivial that $\vert \langle w_s\rangle \rangle$ and $\vert \langle p_s\rangle \rangle$ are the two inputs of the MUL module and the calculation of them has been discussed in Subsection \ref{subsubsec3-3-1}. The SUB module executes the subtraction operation, but before it, a COM module (whose output is $\vert e_0\rangle$) is set in Figure \ref{fig11}. If $\vert e_0\rangle=\vert 1\rangle$ which means the subtrahend is larger than or equal to minuend, i.e. $\vert \langle w_s\rangle\langle p_s\rangle \rangle\geq\vert\langle w_sp_s\rangle\rangle$, and the SET-0 operation is triggered. SET-0 operation could set all input qubits to $\vert 0\rangle$ state by using a sequence of CNOT gates (as used in \cite{chen2018exploring}). In such instances, $\vert e_0\rangle$ becomes a control qubit of the ADD module. It is noteworthy that only when $\vert e_0\rangle=\vert 0\rangle$, the ADD module is activated. After $2^{2n}$ comparisons, SET-0, and addition operations, the final result $\vert\langle r_s\rangle\rangle=\vert\langle w_sp_s\rangle\rangle-\vert\langle w_s\rangle\rangle\vert\langle p_s\rangle\rangle$ is obtained. The state $\vert yx \langle r_s\rangle\rangle$ is akin the quantum representation of ghost image.

\addfigure{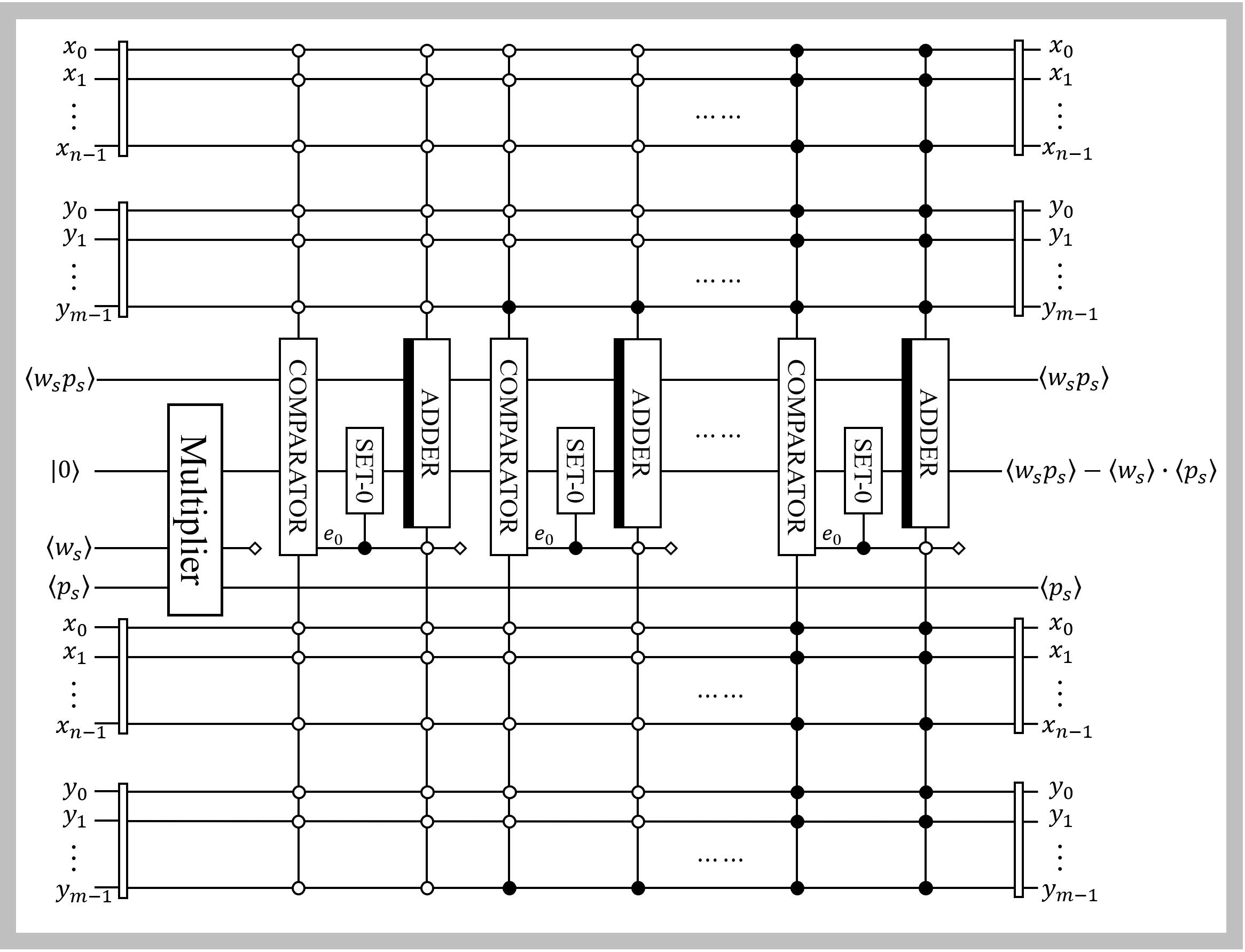}{Circuit realization of QGI protocol $\vert \langle w_sp_s\rangle\rangle-\vert\langle w_s\rangle\rangle\vert \langle p_s\rangle\rangle$}{fig11}{width=0.9\textwidth}

\section{Conclusions}
\label{sec5}
In this study, we proposed a method to implement the ghost imaging experiment by utilizing the quantum circuit. To achieve this, quantum accumulator (ACC), quantum multiplier (MUL), and quantum divider (DIV) modules were proposed. The contributions in this study mainly include: (1) By utilizing the quantum superposition, in quantum register, the capacity to store quantum speckle patterns was enhanced. In addition, by employing the quantum gates (i.e. Hadamard and CNOT gates), a real random speckle pattern sequence was generated, from which the ghost imaging quality was improved by adjusting the sampling rate of the speckle patters. (2) By utilizing the quantum parallelism, some universal quantum arithmetic modules (such as ACC, MUL, and DIV modules) were designed with low computational complexity so that the ghost imaging speed could be assured. Hopefully, such optimized modules would be applicable for other protocols and applications in the quantum computing domain.

Our future work will focus on the following aspects. First, as introduced in the paper, the quantum mask image only includes binary levels for every pixel, i.e $\vert 0\rangle$ or $\vert 1\rangle$, which is a strong astriction for the ghost imaging technique as well as its applications. Therefore, the extension of its chromatic information becomes an important focus in the following studies. Second, it is practical to find a quantum compressive sensing algorithm whose application on the ghost imaging will explore quantum single-pixel imaging via the compressive sampling. Since the ghost imaging usually requires a large number of speckle patterns, such an algorithm may improve the efficiency when a high quality image is obtained in the proposed ghost imaging protocol. Finally, our proposed ghost imaging protocol could be extended to the application of quantum image encryption. Since the inherent properties of ghost imaging, i.e. neither of the two beams in the signal and idler fields carries the information from the object, it is anticipated that the proposed protocol may offer an outstanding performance for the quantum image encryption.


\bibliographystyle{elsarticle-num}


\end{document}